%Paper: hep-lat/9303002
%From: creutz@wind.phy.bnl.gov (Michael J. Creutz)
%Date: Thu, 4 Mar 93 10:24:29 EST

\magnification=1200
% for numbering equations
\newcount\eqnumber
\eqnumber=1
\def\eqnam#1{\xdef#1{\the\eqnumber}}
\def\neweq{\the\eqnumber\global\advance\eqnumber by 1}
% for numbering references
\newcount\refnumber
\refnumber=1
\def\newref{\the\refnumber\global\advance\refnumber by 1}
\def\refname#1{\xdef#1{\the\refnumber}\newref}
\pageno=1
\rightline{IASSNS-HEP-93/11}
\rightline{BNL-xxxxx}
\rightline{March 1993}
\vskip.5in
\centerline{\bf Series expansions without diagrams}
\bigskip
\medskip
\centerline{\bf Gyan Bhanot$^{a,b}$, Michael Creutz$^c$, Ivan Horvath,$^{c,d}$}
\centerline{\bf Jan Lacki$^b$  and John Weckel$^e$}
\bigskip
\bigskip
\baselineskip=18pt

{\it
\vfootnote{$^a$}{Thinking Machines Corporation, 245 First Street,
               Cambridge, MA 02142, USA}
\vfootnote{$^b$}{Institute for Advanced Study, Princeton, NJ 08540, USA}
\vfootnote{$^c$}{Brookhaven National Laboratory, Upton, NY 11973, USA}
\vfootnote{$^d$}{Physics Department, University of Rochester, Rochester,
                 NY, 14627, USA}
\vfootnote{$^e$}{Physics Department, Princeton University, Princeton,
                 NJ, 08544, USA} }

\centerline {ABSTRACT}
\medskip
{\narrower
We discuss the use of recursive enumeration schemes to obtain
low and high temperature series expansions for discrete statistical
systems.  Using linear combinations of generalized helical
lattices, the method is competitive
with diagramatic approaches and is easily generalizable.  We illustrate
the approach using the Ising model and generate low temperature series
in up
to five dimensions and high temperature series in
three dimensions. The method is general and can be applied to
any discrete model. We describe how it would work for Potts models.
}

\vfill\eject
\baselineskip=18pt
{\bf\noindent Introduction}

Expansions about either infinite or vanishing coupling have long been a major
technique for the study of statistical systems
and field theories. These series usually involve a diagrammatic
analysis which becomes rapidly more complex as the order increases.
Thus it would be interesting to have a fully automated technique for the
generation of the relevant terms.

Here we discuss a purely mechanical method to generate
the low and high temperature  expansions for discrete systems.
The approach does not involve explicit graphs, but rather relies on
a recursive computer enumeration of configurations.  We illustrate
the approach on Potts and Ising models, although it is considerably more
general.

The method is based on a recursive transfer matrix procedure of Binder
[\refname\binderref] for the explicit solution of discrete models on small
lattices.  Enting [\refname\entingref] discussed how to combine such
solutions on small lattices to obtain low temperature series.  Guttmann and
Enting have pushed this finite lattice method to obtain rather high order
low temperature series for the three dimensional Ising model
[\refname\guttmannentingref].  Our approach is similar in spirit to this
work, although it differs in many technical details.  In
Ref.~[\refname\bhanotref] these ideas were further developed in the context
of finite size scaling and the analytic structure of the partition
function.  Ref.~[\refname\mcref] explored using these exact counting
procedures on helical lattices to extract the low temperature series.
Helical lattices have been further generalized in Ref. [\refname\letterref]
enabling one to calculate the low temperature series for the three
dimensional Ising model to 50 excited bonds.  Ref.~[\refname\pottsref]
applied these methods to Potts models in two and three dimensions.

This paper is primarily intended to explain these methods in more
detail and explore extensions.
In a recent preprint Vohwinkle [\refname\vohwinkelref]
has adapted the diagrammatic
shadow method to obtain Ising and Potts expansions to several
more terms than we have been able to obtain. Our method is, however,
quite general and easy to implement. It is an open question whether
some of the ideas of Ref. [\vohwinkelref]  can be adapted into
our scheme to get even longer series.

\bigskip
{\bf\noindent Recursive counting}

We begin with a discussion of the recursive approach to solving
small systems exactly.  This section will also serve to establish
notational conventions.
We illustrate the basic method with
the Ising model on a finite three dimensional simple
cubic lattice.  On each site $i$ is a spin $\sigma_i$ taking the
values $\pm 1$.  The energy of the system is
$$
E=\sum_{\{i,j\}}(1-\sigma_i \sigma_j) \eqno(\neweq)
$$
where the sum is over all nearest neighbor pairs of spins, each pair
being counted once. At inverse temperature $\beta$,
the partition function is the sum of the
Boltzmann weight over all configurations
$$
Z=\sum_{\{\sigma\}} e^{-\beta E} \eqno(\neweq)
$$
Organizing the set of configurations by their energy, we rewrite this as a
sum over $E$. This introduces the density of states function $P(E)$
representing the number of
states of the system with the given energy $E$. Thus, we have
 $$ \eqnam\partitioneq
 Z=\sum_{E=0}^{6N}P(E) u^{E/2}\eqno(\neweq)
 $$
where $N$ is the number of sites and $u=e^{-2\beta}$.  If we consider,
for example, an $N^3$ lattice, there will be $2^{N^3}$ states, but the
solution for the partition function can be expressed in terms of
${\cal O}(N^3)$ integers $P(E)$.

For a given lattice we compute the coefficients $P(E)$ exactly using a
transfer matrix to assemble the system one site at a time.  This
recursive construction enables us to build up a lattice with arbitrary
length in one of the three dimensions.  For the series analysis it is
important to always continue the recursion sufficiently to avoid
finite size errors in this ``longitudinal'' direction.  At
intermediate times the process requires an explicit enumeration of any
exposed two dimensional slice.  This effectively reduces the
computational complexity to that of a system of one less dimension.
Thus the solution of an $N^3$ lattice requires, at most, the explicit
enumeration of only $2^{N^2}$ states.  This enables us to work with sizes
which would be impractical for an explicit enumeration of all
states.

The starting point of the method is a list of all states and
corresponding energies for a single transverse layer of the lattice.
All spins outside this layer are frozen to the same value; that is,
the boundary conditions in the longitudinal direction are cold.  Spins
are then sequentially freed to build up the lattice in this third
direction.  At intermediate stages the computer stores the exact
number of states of any given energy and specified exposed top layer.
Storing the top layer as the bits of an integer $I$, define $p(E,I)$
to be this count.  When a new spin or set of spins is added, we obtain
the new counts $p^\prime(E,I)$ as a sum over the old counts
$$\eqnam\recursioneq p^\prime(E,I)=\sum_{I^\prime}
p(E-\Delta(I,I^\prime),I^\prime). \eqno(\neweq) $$ Here $I^\prime$
runs over all integers differing from $I$ at most in the bits
representing the newly covered spins, and $\Delta(I,I^\prime)$ is the
change in energy from any newly changed bonds.  In Ref.~[\mcref] the
spins were added one layer at a time, while here we add them one at a
time.  Thus for the present calculation the sum in the above equation
is only over two terms, representing the two possible values for the
newly covered spin. After the lattice is grown, a sum over the top
layers gives the resulting $P(E)$ for the entire system $$ P(E)=\sum_I
p(E,I). \eqno(\neweq) $$

\bigskip
{\bf\noindent The low temperature series}

Note that as the temperature goes to zero so does the variable $u$.
Thus Eq.~(\partitioneq) is itself the low temperature expansion for $Z$.  From
it, we compute the corresponding series for the average energy per
site,
 $$
\langle E \rangle = {\sum_E E P(E)\over Z}=
2\left(u{\partial \over \partial u}\right) \log(Z) \eqno(\neweq)
 $$
Comparing this expectation value before and after the last spin is added,
we obtain the increase in the average energy per new site.  Expanding this
in powers of $u$ gives
 $$
\langle E/N \rangle = \sum_j e_j u^j\eqno(\neweq)
 $$
We are interested in the coefficients $e_j$ in the infinite volume
limit.

At zero temperature $(\beta=\infty)$ the only states which survive
have all spins parallel.  As the temperature increases, groups of
spins can flip in this uniform background.  A single flipped spin has
6 excited bonds, and thus $e_6$ represents the first non-trivial
term in our expansion.  In our units, each excited bond has energy 2
and there are 6 such bonds for a single flipped spin; thus we have
$e_6=12$.  Continuing to more complicated combinations of flipped
spins gives the usual diagrammatic method to obtain the further coefficients.

Note that any enclosed group of flipped spins must always have an even
number of excited bonds.  Thus the expansion only involves even powers of
$u$.  Our method of construction is such that when a spin is added to
the helix, we account for the energy of both the forward and backward bonds
at once. This, combined with the cold boundary conditions at both ends,
ensures that we generate only even powers of $u$ in our expansions.

\bigskip
{\bf\noindent Helical lattices}

Computing $Z$ exactly on a periodic lattice of size $N\times N\times
N$, the order to which the weak coupling expansion for $\langle E/N
\rangle$ will agree with the infinite volume limit is $4N-2$.  At this
order a line of $N$ flipped spins can wrap around the lattice and show
finite size effects.  Such a configuration will have energy $4N$
rather than the $4N+2$ it would have in infinite space. This is the
smallest excitation affected by boundary effects and hence, on a
periodic lattice of size $N$, the expansion is valid through
${\cal O}(4N-2)$.

The order to which the series is correct can be increased by changing the
boundary conditions to require more spins to be flipped to wrap around the
lattice.  Ref.~[\mcref] showed how a version of helical
boundaries allowed
an $N\times N$ transverse slice to be mimicked with only $[(N^2+1)/2]$
sites.  Here we extend this idea to include the helicity into the direction
in which the lattice is grown.

We build our lattices one site at a time; so, it is natural to imagine
the sites lying in a line.  We do not, however, consider sequential
sites as nearest neighbors.  Instead, we introduce three integer
parameters $\{h_x,h_y,h_z\}$ representing the distance along the line
to the nearest neighbor in the corresponding $x$, $y$, or $z$
direction.  Labeling sites in the sequence by their ordinal number $i$,
the nearest neighbors of site $i$ are at $i\pm h_x$, $i\pm h_y$ and
$i\pm h_z$.  For convenience, we assume
 $$
h_x<h_y<h_z \eqno(\neweq)
$$
With this convention,
as we grow our lattice, all sites more than $h_z$ steps back in the
chain are covered.  Thus the recursive methods of Refs. (1-3) only
require us to keep explicit track of the $h_z$ ``exposed" spins at the
end of our chain.  The computational work grows exponentially with
this number; thus, we wish to keep $h_z$ as small as possible.

A minimal closed loop on such a lattice consists of a number of
steps in each spatial direction such that
 $$\eqnam\loopeq
 n_x h_x+n_y h_y+ n_z h_z=0. \eqno(\neweq)
 $$
Here $n_i$ represents the number of steps in the $i-$th direction.
The length $n$ of such a loop is
 $$
n=\vert n_x\vert+\vert n_y\vert+\vert n_z\vert \eqno(\neweq)
 $$
$n$ is the ``effective'' periodic size of the lattice for our series
construction, and, as argued above, the series is correct to $O(u^{4n-2})$.
On an infinite cubic lattice the only solution  to Eq. (\loopeq) is
the trivial case $n_i=0$.  On a finite lattice, any other solution
represents a finite size correction.  Flipping a chain of spins along
such a closed path generates a state with $4n$ excited bonds, and
creates a potential error in the series at that order. As a simple
example, $(h_x,h_y,h_z)=(3,4,5)$  with
$(n_x,n_y,n_z)=(1,-2,1)$ gives a minimal loop of length 4.  Such
a lattice will give the series equivalent to that on a $4^3$ lattice,
but with only $h_z=5$ sites in cross section.  Similarly,
$(h_x,h_y,h_z)=(19,21,24)$ has closed loops of length 10
corresponding to  $(n_x,n_y,n_z)=(3,-5,2)$.  Here 24 sites mimic
a 10 by 10 cross sectional lattice, thus saving a factor of $2^{76}$
in computational effort.

Note that Eq.~(\loopeq) tells us that if we regard ${\bf n}$ and ${\bf
h}$ as vectors, they are orthogonal.  Thus a simple way to visualize
our lattice is as an infinite one with all sites which lie in any
single plane orthogonal to ${\bf h}$ as identified with each other.
Fig.~(1)
attempts to show this construction.  Considering the plane through the
origin, all the sites lying in this plane themselves form a lattice.
Closed loops that contribute finite size corrections consist of sets
of flipped spins connecting the sites of this lattice.
%The simplest
%such closed loop is labelled A in Fig.~(1) and represents the loop
%closed by lattice periodicity as discussed above.

\bigskip
{\bf\noindent Cancelling loops}

We now discuss how forming linear combinations of the energy series
coefficients from a set of finite helical lattices can give the
infinite volume series to a higher order than any individual lattice
in the set.  The approach here differs in details but is similar in essence
to the combining of partition functions in the finite
lattice method of Refs. [\entingref] and [\guttmannentingref].

Given a set of parameters $(h_x,h_y,h_z)$, it is straightforward to
enumerate the minimal closed paths.  A different set of parameters
corresponds to a different set of such paths.  However, any erroneous
contribution to the coefficients $e_i$ from a particular such path
is, by symmetry, independent of any permutation or sign changes in
the numbers $(n_x,n_y,n_z)$.  This allows us to push the series
further, by combining the results on various size lattices to cancel
the contributions from particular closed loops.

For an explicit example, consider
loops of length 9.
The (16,18,21) lattice has a minimal such loop with steps
${\bf n}=(3,2,-4)$, the $(16,17,21)$ lattice has closed loops with steps
$(1,4,-4)$ and $(5,-1,-3)$, the $(13,18,20)$ lattice has a closed
loops with $(2,3,-4)$ and $(4,-4,1)$, and finally the $(14,17,19)$
system has the loops $(3,2,-4)$ and $(5,-3,-1)$.  If we combine the
coefficients $e_i$ as obtained from these lattices with weights
$(2,1,-1,-1)$ respectively, then all errors from the loops of length
9 cancel out.  This gives the series to the same order as a
lattice with the smallest loop having length 10, which otherwise
would require at least 24 sites.

This procedure extends to cancel further loops.  It is straightforward
to write a program to enumerate the closed loops on various
lattices, and then solve the linear equations to cancel
the errors from such loops.  In Table I we
present a list of 26 lattices and the relative weights for
combining them to cancel all loops of length less than 14.  Note
that in this way we have reduced what would naively require a $14^3$ lattice
to a set of calculations involving a cross section of at most $24$ spins.

After cancelling the single loops as above, a potential problem arises
from more complicated diagrams which wrap around the lattice
simultaneously in two or more ways.  This would correspond to flipping
a set of spins which connects three of the identified sites in
Fig.~(1).
%In Fig.~(1), one such loop is
%shown as B. This loop contributes the next set of finite size
%corrections after loops of type A.
In selecting our lattices for
Table I we did not consider
any system which had a loop
%such as B
contributing to any order for
which we were extracting the series coefficient.

It is easy to calculate the order at which these more complex loops
%such as B
contribute.
In our lattice finding program we first find the three closest identified
sites which do not lie on a single straight line.
(Double loops connecting points in a line are automatically
cancelled at the same time as the simple loops.)
%A types of loops.
Denote the minimum distances between these three images
as $d1$, $d2$, and $d3$.  In most cases, the minimal way of
flipping a set of spins to wrap around
these three loops produces an energy of $(d-1)\times(d1+d2+d3-2)-2$,
where $d$ is the dimension of the system.  We rejected using any lattices
for which this number is at or below the order to which we were extracting
the series.

In rare special cases this formula needs a correction.
The energy can be lower
if one of the fundamental loops has no steps in one direction.  Then
the two loop diagram can run into its periodic image, reducing the relevant
order.  For example, with the ${\bf h}=(11,15,18)$ lattice, the fundamental
loops have ${\bf n}=(0,6,-5)$, $(3,-1,-1)$ and $(3,5,-6)$.  The minimal energy
for a set of flipped spins which connects these three images is 52 bonds,
rather than the predicted 54 from the above formula.
Needless to say, this lattice caused us considerable consternation.

The utility of these cancellations depends strongly on dimension.
For two dimensions with at most $h_y$ sites on the top row, the best
solution is always a single lattice with $h_x=h_y-1$.  In this case the
shortest extraneous loop has ${\bf n}=(h_y,-h_y+1)$ with length $2h_y-1$.
Note that as $h_y$ becomes large,
the transfer matrix effectively grows the lattice along a diagonal.

For higher dimensions, on the other hand, there are a rapidly growing
number of interesting lattices to cancel loops between, and this method
becomes particularly powerful.  Table II includes a list of 15
lattices which give the four dimensional series through order 50 excited bonds.
Although the largest lattice here has 28 sites in the top row, the tricks
of the next section are also more effective in four dimensions, so this
is not a particularly difficult case.

Note that although we have been discussing these lattices in the context
of the Ising model, the results are more general.  In particular,
the combinations in Table I are valid for any nearest neighbor
model on a simple cubic lattice.

These methods can also be applied to other than simple cubic lattices.
For example, to treat a body centered cubic lattice, each site has
eight neighbors, so we need four components for $h$. We can merely use a four
dimensional lattice-finding program modified to require the real
closed loops of length 3 be present and not be cancelled.

\midinsert
$${\baselineskip=12pt
 \vbox
{\noindent
Table I.  A combination of 26 lattices which gives the three dimensional
low temperature
expansion coefficients through 54 excited bonds.  The first column represents
the coefficient with which the lattice is to be weighted and the second
gives the vector ${\bf h}$ which defines the lattice.
\medskip
\settabs 2 \columns
\+ coefficient & $(h_x,h_y,h_z)$ \cr
\smallskip
\+ 1  & (17,23,24)  \cr
\+ 2  & (19,22,24)  \cr
\+ -1  & (19,21,24)  \cr
\+  3 & (19,20,23)  \cr
\+  -1 & (18,20,23)  \cr
\+  -4 & (11,15,23)  \cr
\+  -1 & (18,21,22)  \cr
\+  3 & (16,21,22)  \cr
\+ -2  & (18,19,22)  \cr
\+ -3  & (15,19,22)  \cr
\+ 3  & (14,19,22)  \cr
\+ -3   & (16,17,22)  \cr
\+  -3 & (5,18,21)  \cr
\+ 3   & (8,17,21)  \cr
\+ 3  & (7,19,20)  \cr
\+  -6 & (1,17,20)  \cr
\+ 6  & (16,18,19)  \cr
\+ -2  & (16,17,19)  \cr
\+ 6  & (12,17,19)  \cr
\+ 7  & (8,17,18)  \cr
\+ -3  & (7,16,18)  \cr
\+ -9  & (11,14,18)  \cr
\+ -7  & (8,13,18)  \cr
\+ 2  & (9,16,17)  \cr
\+ 3  & (1,13,15)  \cr
\+ 4  & (12,13,14)  \cr
\medskip
}
}
$$\endinsert

\midinsert
$${\baselineskip=12pt
 \vbox
{\noindent
Table II.  A combination of 15 lattices which gives the four dimensional
low temperature
expansion coefficients through 50 excited bonds.  The first column represents
the coefficient with which the lattice is to be weighted and the second
gives the vector ${\bf h}$ which defines the lattice.
\medskip
\settabs 2 \columns
\+ coefficient & $(h_x,h_y,h_z)$ \cr
\smallskip
\+ 3   &    (15,24,25,28)  \cr
\+ -27   &    (15,21,25,28)  \cr
\+   14 &    (13,20,25,28)  \cr
\+  27  &    (15,20,26,27)  \cr
\+  16  &   (11,20,26,27)  \cr
\+  18 &    (19,20,25,27)  \cr
\+ 2  &    (11,15,25,27)  \cr
\+ -4    &    (16,17,23,27)  \cr
\+ -13   &    (14,17,19,27)  \cr
\+ -4   &    (11,20,25,26)  \cr
\+ -6   &    (15,18,25,26)  \cr
\+ -16  &    (15,21,23,26)  \cr
\+ -16   &    (7,20,24,25)  \cr
\+ 28   &    (14,15,23,25)  \cr
\+ -21   &    (17,18,22,25)  \cr
\medskip
}
}
$$\endinsert

\bigskip
{\bf\noindent Miscellaneous tricks}

During the recursive buildup of the lattice, each new count is the sum of
just two terms, representing the two possibilities for the covered spin.
Thus the arithmetic involved is rather trivial.  On the other hand, we must
store counts for all energies up to the maximum order desired as well as
for all relevant values of the top $h_z$ spins of our helical lattice.  In
addition, the intermediate counts can become rather large numbers.  Thus,
the primary computational problem is storage.  To substantially reduce
these demands, we calculated the series coefficients several times, each
time modulo a small different integer.  Depending on the integers chosen,
this enabled us at intermediate stages to store the counts in either one
byte or one short integer each.  As all operations are simple additions or
multiplications, this procedure correctly gives the final coefficients
modulo the given integers.  After multiple passes using mutually prime
values for these modulos, we use the Chinese remainder theorem to
reconstruct the final series.  This theorem states that if you know a
number modulo a set of relatively prime integers, then the number is
uniquely determined up to the product of those integers.

As we are repeating the series calculations for several different
modulos and for several different lattices and only combining the
results at the end, this problem is particularly suitable for
trivial parallelization.  Indeed, except for the most memory intensive
cases, we have experimented quite successfully
with sending different lattice-modulo combinations to a farm of
workstations.  For this we have been using the Condor distributed
batch system [\refname\condorref].

Note that as we add spins, the energy of the system can only increase.
This means that we never need counts involving more excited bonds than the
order to which we are evaluating the series.  Furthermore,
while the recursive procedure is predicated on keeping all top rows
for the lattice, this is not actually necessary if we only want the
series to some given order.  In particular, we need not store any counts
for top rows which already contain more excited bonds than the
order we are working to.  To handle this, we use a simple subroutine that,
given a possible top row, finds the next top row in numerical sequence
with a number of excited bonds less than or equal to the working order.

As an explicit example with the Ising model,
consider the ${\bf h}=(17,23,24)$ lattice and allowing
only up to 54 excited bonds.  In this case
we need keep only 2,778,176 of the possible
$2^{24}=16,777,216$ possible top rows.
In four dimensions, because there are
additional bonds which can be excited, the corresponding reductions are
even larger.

In addition to not keeping all top rows, we need not store counts with
less energy than the minimum possible for a given top row.  That is,
while for the top row with all spins up we need to keep counts for all
possible excitation energies up to the order under consideration, if
the top row has a single flipped spin we need only keep counts of at
least 3 excited bond pairs, and so on.  Finally, for the Ising case on
a simple cubic lattice with our boundary conditions there can only be
an even number of excited bonds.  In this way, the above (17,23,24)
lattice requires keeping track of $11,259,428$ individual $p_0(E,I)$,
or less than one count per possible top row.

During the recursion, each new count is the sum of one or two of the
old ones, corresponding to whether the covered spin is flipped or not, and
whether for a flipped spin we do not already have more energy than being
considered for the count in question.  A simple way to implement
this is to have two index arrays, with the elements of each representing
the location of the old counts to be used.  Having an entry
in the index array out of bounds provides
a simple way to flag those cases where only a single term goes into
the sum.  Once the geometry is established by the construction of
these arrays, the program simply
loops over the counts, making the new values the sum of two old ones
pointed to by these indices.  In this way all the complications of
setting up the geometry need only be done once per lattice.

One can save additional memory by not storing the full indices,
but using the fact that if one orders the counts first by top row numerically,
and then by bonds, the respective indices always change by relatively
small numbers in going from one count to the next.  Thus we need only store
the changes rather than the indices themselves.  In the main loop
the new indices are obtained
by a simple addition to the previous ones.  The index changes for our studies
could all be stored in a single byte.

A final trick that we have so far only used minimally is to invert
a partially grown lattice on itself.  The idea is that given
the counts for all possible top layers, we can then obtain the
counts for a lattice roughtly twice as long with all possible specified
layers in the middle.  Calling this count $p_d(E,I)$, we have
$$
p_d(E,I)=\sum_{E_1,E_2} p(E_1,I) p(E_2,I_r) \delta(E,E_1+E_2-d(I))
  \eqno(\neweq)
$$
where $I_r$ has the bits of $I$ in reversed order (because the lattice
has been flipped upside down)  and $d(I)$ represents
the excited bonds inside the middle layer.  The latter is removed to
prevent double counting.
This technique can provide information on
correlation functions in this middle layer.  As
all states are known explicitly, any such correlation function can
be obtained exactly with no significant additional drains on computer
time or memory.  As a simple example, this
provides an alternative method for
obtaining the magnetization series to that discussed in the next
section.

\bigskip
{\bf\noindent Other observables}

So far we have been discussing the direct low temperature expansion
for the partition function or,
equivalently, the average energy or the specific heat.
The method easily extends to other
observables by generalizing the counts.  For example, consider applying
a magnetic field by generalizing the partition function to
 $$
Z=\sum_{\{\sigma\}} e^{-\beta E-H\sum_i \sigma_i} \eqno(\neweq)
$$

Derivatives with respect to the applied field give us a procedure to
compute the
magnetization
$$
M=\langle \sigma_i \rangle=-{1\over N} {\partial \over {\partial H}}
  \log Z \eqno(\neweq)
$$
and the magnetic susceptibility
$$
\chi={\partial \over \partial H} M  \eqno(\neweq)
$$

For general $H$ one can expand observables simultaneously in $u$ and
$\lambda=e^{-2H}.$  In Ref. [\letterref] this possibility was discussed in
terms of generalizing the counts $P(E)$ to the two indexed count $P(E,S)$,
representing the number of states of a given bond energy and number of
flipped spins $S$.
The recursion relations for these counts are completely analagous to
those for $P(E)$.  The double series for the magnetization was presented
up to order 42 excited bonds in Ref. [\letterref].

One difficulty with this approach is the increased memory required for
storing counts for all magnetizations as well as energies.  If one is only inte
rested in the magnetic properties in the zero field limit, one
can store considerably less.
In particular, consider moments of the magnetization, from which
quantities such as the susceptibility are easily extracted.  It is
convenient to define new quantities
$$
P_k(E)=\sum_S S^k P(E,S).  \eqno(\neweq)
$$
With this definition, $P_0(E)$ is simply the original count $P(E)$.  The
zero field magnetization is easily found from
$$
M=1-2{\sum_E P_1(E) e^{-\beta E} \over N Z}.  \eqno(\neweq)
$$
Finally, from $P_2$ we can obtain the magnetic susceptibility
$$
\chi= 4\left({\sum_E (P_2(E)-P_1(E)^2) e^{-\beta E} \over N Z}\right).
 \eqno(\neweq)
$$

The advantage of working with these moments is that they themselves satisfy
simple recursion relations.  To derive them, consider the generalization
of Eq.~(\recursioneq)
 $$\eqnam\momenteq
 p^\prime(E,S,I)=\sum_{I^\prime}
 p(E-\Delta(I,I^\prime),S-\Delta_s(I'),I^\prime). \eqno(\neweq)
 $$
Here $p(E,S,I)$ is the number of states of energy $E$, with $S$ flipped
spins, and with lattice top row specified by $I$, and  $p^\prime$ is the same
quantity on the new lattice obtained after adding the new spin.
We denote by $\Delta_s(I')$ the change in the number of flipped
spins; that is, $\Delta_s=1$ if the new spin is flipped (the
relevant bit of $I'=1$) and $\Delta_s=0$ otherwise.

Now define the moments,
$$
p_k(E,I)=\sum_S S^k p(E,S,I).  \eqno(\neweq)
$$
Taking moments of Eq.~(\momenteq) now gives the recursion relations for
the $p_k$
$$\eqalignno
{ p_0^\prime(E,I)&=\sum_{I^\prime}
 p_0(E-\Delta,I^\prime) &(\neweq)\cr
 p_1^\prime(E,I)&=\sum_{I^\prime}
   p_1(E-\Delta,I^\prime)
   +\Delta_s(I)p_0(E-\Delta,I^\prime) &(\neweq)\cr
 p_2^\prime(E,I)&=\sum_{I^\prime}
   p_2(E-\Delta,I^\prime)
   +2\Delta_s(I)p_1(E-\Delta,I^\prime)
   +\Delta_s^2(I)p_0(E-\Delta,I^\prime). &(\neweq)\cr
}
$$
The first of these relations is just our original recursion,
and the others enable us to calculate the magnetization and
susceptibility with the addition of only two new counts.

It is straightforward to derive the analogous counting
schemes for n-point susceptibilities and their various spatial
moments, like the second moment of 2-point susceptibility
$\mu_2=<x^2 \sigma_x \sigma_0>$. In later case however
there are some conceptual
difficulties connected to the ambiguity of the definition
of the coordinate on the helical lattice. Some more work needs
to be devoted to this problem.

\bigskip
{\bf\noindent Strong coupling}

We now turn to the application of the counting methods to the
strong coupling series.  In this section we describe the procedure
for the three dimensional Ising model, although again it is easily
generalized.  As before we consider spins $S_i$
on the lattice sites $i$ and taking the values $\pm 1$.  The partition function
of Eq.(2) can be trivially rewritten
$$
Z=({{1+e^{-2\beta}}\over{2}})^{N_l}\sum_\sigma \prod_{\{i,j\}}
 (1+\sigma_i\sigma_j\tanh(\beta))
\eqno(\neweq)
$$
where the product is over all lattice links and
$N_l$ is the number of links in the system.  For the strong coupling
series we consider small $\beta$ and expand the above sum in powers of
$\tanh(\beta)$.  Each term involves a set of selected bonds which each give
a power of $\tanh(\beta)$.  Having selected a set of bonds, we can then
perform the sum over the spins.  If any site has an odd number of
selected bonds eminating from it, the sum will vanish.  Otherwise the
sum over any given spin gives a factor of two.  Thus we conclude
$$
Z=2^{N}({{1+e^{-2\beta}}\over{2}})^{N_l}\sum_k N(k) (\tanh(\beta))^k
\eqno(\neweq)
$$
Here N(k) represents the number of possible ways to select $k$ bonds
in such a manner that each site is the end of an even number of selected
bonds.  We adapt our counting methods to evaluate these numbers $N(k)$.

As before we maintain information on the top layer of our lattice while
adding new sites one at a time.  Here, however, rather than the values
of the spins themselves on the top layer, we keep information on the
selected bonds ending there.  In particular, because we want to allow
future bonds to extend above the top row, we relax the
constraint that an even
number of bonds end on the top sites.  Thus, we keep a count $N(k,I)$
where $I$ now stores in its set bits those sites with an odd number of
bonds coming into them from previous sites.  We refer to sites
with an odd number of incoming bonds as having ``loose ends''  or
``dangling bonds.''
On adding a new site, we have the basic recursion relation
$$
N^\prime(k,I)=\sum_{I^\prime} N(k-\Delta(I,I^\prime), I^\prime)
\eqno(\neweq)
$$
where $\Delta(I,I^\prime)$  represents the number of selected bonds
attached to the new spin and $I^\prime$ is related to $I$ via changes
in those bits representing sites attached to the new one.

In three dimensions, for any given $(k,I)$ there will be four terms
in the above sum over $I^\prime$.
This represents a factor of two for whether the new $x$ bond is selected
times a factor of two for whether the new $y$ bond is chosen.  Whether
the corresponding $z$ bond is chosen or not is determined by the
corresponding bit of $I$ which determines if an even or odd number
of bonds are selected.

An immediate factor of two in memory is saved because each bond
has two ends.  This means that if no bonds enter from outside
below the lattice,
the top layer must have an even number of loose ends.  Any top layers
with an odd number of loose ends need never be kept.
In practice, instead of looping over all given any integers $I$
representing the dangling
bonds from an allowed configuration, we need only loop over the
right $h_z-1$ bits of $I$ and can determine the allowed leftmost bit
by parity considerations.

We work with generalized helical lattices as before.  For simplicity
in initialization, we set all counts to zero except for $I=0$, representing
no dangling bonds.  This may seem a bit peculiar because we do not allow loops
to enter and travel through the bottom layer.  It is, however, simple
to implement and boundary conditions in the longitudinal direction
are irrelevant if we grow the lattice long enough.

On a single helical lattice,
the strong coupling series will be correct to the order of the first chain
of bonds which wraps around one of the artificial closed loops
discussed earlier.  The double loop criterion is somewhat different now;
here it is only the total length of a loop which wraps around
two directions that matters.
Rejecting lattices with such double loops, we can perform the same cancellation
between lattices as in the low temperature series.

The strong coupling series can be extended significantly by using
the fact that all valid loops of links on an infinite lattice will
have an even number of selected bonds in any of the coordinate directions.
We use this fact by calculating the counts several times, but including
extra minus signs when adding bonds in various directions.  For
example, if we first find the series giving every $x$ bond a weight of -1,
we can then add the result without this extra sign and any artificial
diagram involving
an odd number of $x$ bonds will cancel out.  Thus we need not worry
about any finite size effects involving an odd number of steps
in the $x$ direction.  Repeating the procedure 8 times for all combinations
of minus signs for the three possible directions, we can ignore any extraneous
closed loops with an odd length along any dimension.
Similarly, any double loops with an odd number of steps in any direction
can also be ignored.  Without this trick the order to which the strong coupling
series can be found is rather uninteresting.

\midinsert
$${\baselineskip=12pt
 \vbox
{\noindent
Table III.  A combination of 11 lattices which gives the three dimensional
strong coupling series through order 20.  The first column represents
the coefficient with which the lattice is to be weighted and the second
gives the vector ${\bf h}$ which defines the lattice.
\medskip
\settabs 2 \columns
\+ coefficient & $(h_x,h_y,h_z)$ \cr
\smallskip
\+  4 &  (4,15,16)  \cr
\+  2 &  (12,13,16)  \cr
\+ -4 &  (4,13,16)  \cr
\+ -6 &  (11,12,16)  \cr
\+  4 &  (7,12,16)  \cr
\+  6 &  (11,14,15)  \cr
\+ -4 &  (7,13,15)  \cr
\+  3 &  (5,11,15)  \cr
\+  3 &  (9,13,14)  \cr
\+ -3 &  (4,12,13)  \cr
\+ -3 &  (8,11,13)  \cr
\medskip
}
}
$$\endinsert

\midinsert
$${\baselineskip=12pt
 \vbox
{\noindent
Table IV.  The coefficients for the
strong coupling series for the three dimensional Ising model
through order 20.  The $f_k$ are defined in the text.
\medskip
\settabs 2 \columns
\+  k & $k f_k$ \cr
\smallskip
\+  0   &    0   \cr
\+  2   &    0   \cr
\+  4   &    12   \cr
\+  6   &    132   \cr
\+  8   &    1,500   \cr
\+  10   &   19,800   \cr
\+  12   &   288,528   \cr
\+  14  &    4,468,380   \cr
\+  16  &    72,236,124   \cr
\+  18  &    1,206,062,448   \cr
\+  20  &    20,649,134,532   \cr
\medskip
}
}
$$\endinsert

With these tricks, we have found the series through 20 selected
bonds from the combination of lattices given in Table (III).  As the
lattice size goes to infinity, we write the
free energy in the form
$$
F= {\log(Z) \over N_S}=\log 2+{N_L\over N_S}\log({{1+e^{-2\beta}}\over 2})
+\sum_k f_k (\tanh(\beta))^k
\eqno(\neweq)
$$
where $N_S$ and $N_L$ denote the number of sites and links, respectively.
To extract the coefficients $f_k$, it is somewhat easier to work
with the analog of an expectation,
$$
\langle k \rangle= {\sum_k k N(k) \over \sum_k N(k)}
\longrightarrow N \sum_k k f_k
\eqno(\neweq)
$$
As for the low temperature series, we extract the contribution per spin
by comparing the counts before and after adding the last spin.  Since they
are just combinations of integer counts,
the products $kf_k$ themselves are always integers, while the $f_k$
are not in general.  We tabulated  these numbers through order
20 in Table (IV).  These numbers are not new; for example they plus one
additional term follow from the results in Ref.~[\guttmannentingref].

\bigskip
{\bf\noindent Potts models}
\medskip

As we mentioned earlier, the application of the counting techniques
to the low temperature series expansions is
very easily generalizable to any discrete system with nearest neighbor
interaction. To illustrate this, consider the q-state Potts
model, defined by the interaction of the form
$$
E=\sum_{\langle ij \rangle}\left[1-\delta_{\sigma _i,\sigma _j}
\right]  \eqno(\neweq)
$$
where $\sigma _i$ is a site-defined field that takes $q$ possible
values. The sum is taken over all nearest neighbor pairs of spins
with $\delta$ being the Kronecker symbol.

Writing the partition function in the form
$$
Z=\sum_{E=0}^{dN}P(E) u^{E} \eqno(\neweq)
$$
with $d$ being the spatial dimension and $u=e^{-\beta}$, one can follow
essentially the same steps we outlined in the discussion of the
Ising model. Namely, the application of recursive counting using
helical lattices and Chinese arithmetic comes through with no
change at all. The differences are of a technical nature only, the
conceptual ones.

Working with $h$ spins on a helix, the maximum number of configurations of
the top layer is $q^h$. Since a single bit is no longer sufficient to keep
the state of the individual spin, it would be more complicated
to code the state of
the top layer in a single word.  Instead, we use several words to
represent each top layer configuration.  For example, for the $q=3$
calculations we used two words per configuraion while for $q=8$ three words
were required.  It is also clear that now the analogue of Eq.~(4) has $q$
terms, corresponding to the $q$ different possible values of the added
spin.

We have computed the low temperature expansions for the energy,
magnetization and susceptibility for the $q=3$ model in $d=2$ and $d=3$ and
the $q=8$ model in $d=2$.  The resulting series have been extensively
discussed and analyzed in [\pottsref] and we do not repeat them here.

\bigskip

{\bf\noindent Results and analysis}
\medskip

Using these methods with the simple Ising model,
we obtained the series for the average energy per bond given in Table
(V).  In Table (VI) we give the series through order 54 excited
bonds for the magnetization and the magnetic susceptibility of
the three dimensional model.

\midinsert
{\baselineskip=12pt
\vbox
{\noindent
Table V.  The low temperature expansion coefficients for the average
energy per unit volume for the three, four, and five dimensional Ising model
on a simple cubic lattice.
\medskip
\settabs 4 \columns
\+ $i$  & $e_i$ (3-d) & $e_i$ (4-d)& $e_i$ (5-d)\cr
\smallskip
\+ 0 &  0  & 0 & 0 \cr
\+ 2 &  0  & 0 & 0 \cr
\+ 4 &  0  & 0 & 0 \cr
\+ 6 &  12  & 0 & 0  \cr
\+ 8 &  0   & 16  & 0 \cr
\+ 10 &  60  & 0 &  20 \cr
\+ 12 &  -84  & 0 & 0  \cr
\+ 14 &  420  & 112 &  0 \cr
\+ 16 &  -1,056  & -144 &  0 \cr
\+ 18 &  3,756  & 0 &  180 \cr
\+ 20 &  -11,220  & 1,120 & -220  \cr
\+ 22 &  37,356   & -2,816 & 0  \cr
\+ 24 &  -118,164  & 2,032 &  0 \cr
\+ 26 &  389,220  & 11,856 &  2,340 \cr
\+ 28 &  -1,261,932  & -46,704 & -5,600  \cr
\+ 30 &  4,163,592   & 66,960 &  3,320 \cr
\+ 32 &  -13,680,288  & 94,576 & 640  \cr
\+ 34 &  45,339,000   & -707,472 & 32,980  \cr
\+ 36 &  -150,244,860  & 1,545,120 &  -122,220 \cr
\+ 38 &  500,333,916   & -148,656 & 145,540  \cr
\+ 40 &  -1,668,189,060  & -9,522,864 & -31,420  \cr
\+ 42 &  5,579,763,432   & 30,130,576 & 454,860  \cr
\+ 44 &  -18,692,075,820  & -30,299,808 & -2,483,360  \cr
\+ 46 &  62,762,602,860   &  -104,198,096 & 4,560,440  \cr
\+ 48 &  -211,062,133,044  & 520,429,776 & -2,922,240  \cr
\+ 50 &  711,052,107,060  & -918,744,400 &  6,717,220 \cr
\+ 52 &  -2,398,859,016,684 &  &   \cr
\+ 54 &  8,104,930,537,260 &  &   \cr
}
}
\endinsert

\midinsert
{\baselineskip=12pt
\vbox
{\noindent
Table VI.  The low temperature expansion coefficients for the average
magnetization and magnetic susceptibility for the three dimensional Ising model
on a simple cubic lattice.
\medskip
\settabs 3 \columns
\+ $i$  & $M_i$ & $\chi_i$ \cr
\smallskip

\+ 0  &       1  &   0   \cr
\+ 2 & 0 & 0  \cr
\+ 4 & 0 & 0  \cr
\+ 6 & -2 & 1  \cr
\+ 8 & 0 & 0  \cr
\+ 10 &   -12 & 12  \cr
\+ 12 & 14 & -14  \cr
\+ 14 & -90 & 135  \cr
\+ 16 & 192 & -276  \cr
\+ 18 & -792 & 1,520  \cr
\+ 20 & 2,148 & -4,056  \cr
\+ 22 &   -7,716 & 17,778  \cr
\+ 24 & 23,262 & -54,392  \cr
\+ 26 & -79,512 & 213,522  \cr
\+ 28 & 252,054 & -700,362  \cr
\+ 30 &   -846,628 & 2,601,674  \cr
\+ 32 & 2,753,520 & -8,836,812  \cr
\+ 34 & -9,205,800 & 31,925,046  \cr
\+ 36 & 30,371,124 & -110,323,056  \cr
\+ 38 & -101,585,544 & 393,008,712  \cr
\+ 40 & 338,095,596 & -1,369,533,048  \cr
\+ 42 &   -1,133,491,188 & 4,844,047,090  \cr
\+ 44 &   3,794,908,752 & -16,947,396,000  \cr
\+ 46 &   -12,758,932,158 & 59,723,296,431  \cr
\+ 48 & 42,903,505,030 & -209,328,634,116  \cr
\+ 50 &   -144,655,483,440 & 736,260,986,208  \cr
\+ 52 & 488,092,130,664 & -2,582,605,180,212  \cr
\+ 54 &   -1,650,000,819,068 & 9,074,182,912,884  \cr
}
}
\endinsert

We will now give a brief analysis of our series to get results
on the critical temperature and exponents. In the usual Dlog Pade (DlP)
analysis [\refname{\paderef}], given a series expansion for $F(u)$ to
$N$-th order,
$F_N(u)=1+\sum_{i=1}^N f_iu^i$, (we will use the simplification that one
can always normalize the series so that the constant term is unity),
one computes coefficients for polynomials $Q_L(u)=\sum_{i=0}^L
q_iu^i$ and $R_M(u)=1+\sum_{i=1}^M r_iu^i$, which satisfy,
$$
Q_L(u)/R_M(u) = F'_N(u)/F_N(u)\eqno(\neweq)
$$
to $O(u^N)$ with $L+M=N-1$.

The position $u_c$ of a singularity of $F$ of the form $F\sim
A/|u-u_c|^\zeta$ will be approximated by the zeros of $R_M$. The
exponent $\zeta$ is estimated by $\zeta = -Q_L(u_c)/R'_M(u_c)$.

In addition to DlP, we will also use the method of inhomogeneous
differential approximants (IDA) introduced by Fisher and Au-Yang
[\refname{\fisheryangref}] (see also [\refname{\hunterbakerref}]).
These are
useful in handling singularities of the form,
$$
F(u)={{A(u)}\over{|u-u_c|^{\zeta}}} +B(u)\eqno(\neweq)
$$ where $A$ and $B$ are analytic in $u$.
In this method, one computes coefficients for polynomials
$Q_L(u)$, $R_M(u)$ and $S_J(u)$ which satisfy,
$$
F_N Q_L +S_J =F'_N R_M\eqno(\neweq)
$$
to order $N$, with $L+M+J=N-2$.  Here the subscripts represent the
highest order present in the given polynomials.

Note that for $S_J=0$ one gets the usual Dlog Pade ratio from
$Q_L/R_M$.  It is easy to see that potential critical points $u_c$ are
the zeros of $R_M$ and for each of these, the exponent $\zeta$ is again
estimated as $\zeta=-Q_L(u_c)/R'_M(u_c)$.

% First consider the weak coupling results.
In Fig.~2 we show the critical
point $u_c$ and the exponent $\beta$ obtained from the magnetization series
in Table (VI) using DlP and IDA analysis.  The different points represent
different values for the orders of the various polynomials used in the
analysis.  The critical value $u_c$ is most accurately known from Monte
Carlo calculations and is shown as a vertical line. The thickness of this
line represents the accuracy with which this number is known.  As is clear,
the DlP gives a $u_c$ quite far from the Monte Carlo value and the IDA
results tend to scatter around it. If however, we take $u_c$ as given, we
can calculate $\beta$ quite accurately from our results by interpolating
the IDA results to the known $u_c$.  Fitting a straight line to the points
near $u_c$ in Fig.~(2) gives $\beta=0.289(1)$ where the errors are only the
errors on the fit and ignore possible systematic effects from the unknown
higher order terms in the series.

Similarly, the susceptibility series gives the results in Fig.~3. Now
the results for the DlP and the IDA are mixed in together, in contrast
to the magnetization series results (see above).
Once again, the critical point is more
accurately obtained from Monte Carlo data, and, assuming that value, we find
the exponent $\gamma=1.281(1)$ from straight line fits
to the data in the critical region.

Finally, we turn to the heat capacity, $C_v$, series. The results are
shown in Fig.~(4).
The DlP lie deceptively on a straight line which, if extrapolated to
the Monte Carlo value for $u_c$ gives the wrong answer near
$\alpha=0.2$ as was already noticed in our earlier paper
[\refname{\letterref}]. The IDA
results on the other hand, are steeply varying in the region of $u_c$.
If we take the three points from the $J=0$ IDA data and fit them to a
line, we get $\alpha=0.128(1)$.
\bigskip

{\bf\noindent Concluding remarks}

The method presented here should easily generalize to other discrete
systems.  The helical lattices used, as well as the combinations to
cancel out finite size errors, are independent of the specific model.
It is straightforward to introduce additional couplings, although
this will increase memory needs.  Some interesting possibilities for
futher exploration are gauge and coupled gauge-spin
models in various dimensions.  Changing boundary conditions should
enable the study of interface properties.
In Ref.~[\refname\fermionref]
similar recursive
methods were suggested as a means to study many fermion systems.  A
particularly challenging problem is the extension of these ideas to
theories with continuous spins.  Some work along these lines for gauge theories
appears in Ref.~[\refname\narayananvranasref]

\vskip .5in
{\bf Acknowledgements}
\vskip .2in
We thank Joseph Straley for discussions on turning finite lattice
partition functions into low temperature series. We also thank David
Atwood for discussions on the Chinese remainder theorem as a memory
saving trick.
We thank T. Guttmann for a preliminary copy of Ref.~[\guttmannentingref].
We thank Professor M. E. Fisher for many useful comments.
Some of the computations were done on the Cray-YMP at the
Supercomputing Computations Research Institue at Florida State University,
and some were done on the Cray computers at the National Energy Reserch
Supercomputer Center, Livermore, CA.  The order 54 three dimensional
Ising calculations were completed on Connection Machines at Thinking
Machines Corporation.
This work has been authored under Contracts No. DE-AC02-76CH00016 and
DE-FG02-90ER40542 of the U.S.~Department of Energy.  Accordingly, the
U.S.~Government retains a nonexclusive, royalty-free license to publish or
reproduce the published form of this contribution, or allow others to do
so, for U.S. Government purposes. The work of GB was also partly supported
by a grant from the Ambrose Monell Foundation.  The research of JL was
supported by the Swiss National Scientific Fund.

\vskip .5in
{\bf\noindent Figure captions}
\vskip .2in
Fig. 1.  Visualizing the helical lattice.  All lattice points lying
in the plane orthogonal to the vector ${\bf h}$ are to be identified.

Fig. 2.  The critical coupling and exponent $\beta$ obtained from the
magnetization series for the three dimensional Ising model.

Fig. 3.  The critical coupling and exponent $\gamma$ obtained from the
magnetic susceptibility series for the three dimensional Ising model.

Fig. 4.  The critical coupling and exponent $\alpha$ obtained from the
energy series for the three dimensional Ising model.

\vfill\eject
{\bf \noindent References}
\vskip .2in
\item{\binderref.} K.~Binder, Physica 62 (1972) 508.

\item{\entingref.} I.G.~Enting, Aust.~J.~Phys. 31 (1978) 515.

\item{\guttmannentingref.} A.J.~ Guttmann and I.G.~Enting, J.~Phys.~A
(submitted).

\item{\bhanotref.} G.~Bhanot, J.~Stat.~Phys. 60 (1990) 55; G.~Bhanot and
S. Sastry, J. Stat. Phys. 60 (1990) 333.

\item{\mcref.} M.~Creutz, Phys.~Rev.~B43 (1991) 10659.

\item{\letterref.} G.~Bhanot, M.~Creutz, and J.~Lacki, Phys.~Rev.~Letters 69
(1992) 1841.

\item{\pottsref.} G. Bhanot, M. Creutz, U. Gl\"assner, I. Horvath,
J. Lacki, K. Schilling, and J. Weckel, preprint (1993).

\item{\vohwinkelref.} C. Vohwinkel, preprint (1992).

\item{\condorref.} M. Litzkow, M. Livny, and M. Mutka, Proc. 8th Int. Conf.
 on Distributing Computing Systems, San Jose, 1988.

\item{\paderef.} G.A.~Baker, {\it Essentials of Pad\`e Approximants} (Academic
Press, New York, 1975).

\item{\fisheryangref.}  M.E.~Fisher and H.~Au-Yang, J. Phys. A12 (1979) 1677.

\item{\hunterbakerref.} D.L.~Hunter and G.A.~Baker, Phys. Rev. B19 (1979)
3808.

\item{\fermionref.} M.~Creutz, Phys.~Rev.~B45 (1992) 4650.

\item{\narayananvranasref.} R. Narayanan and P. Vranas, Nucl. Phys. B330
(1990) 608.

\vfill\eject
\bye